\documentclass[final,3p,times]{elsarticle}
\usepackage{graphicx,amssymb,amsmath,subfigure,lineno,natbib}
\usepackage{framed}
\usepackage{epsfig,mathptmx,times,color,flushend,gensymb,fancyhdr}
\usepackage{soul}
\definecolor{shadecolor}{rgb}{1,0.8,0.3}

\UseRawInputEncoding
\begin{document}
 %\linenumbers
\begin{frontmatter}

\title{Towards Agrobots: Identification of the Yaw Dynamics and Trajectory Tracking of an Autonomous Tractor}

\author[KULeuven]{Erkan Kayacan}
\author[NTU]{Erdal Kayacan}
\author[KULeuven]{Herman Ramon}
\author[KULeuven]{Wouter Saeys}
\address[KULeuven]{Department of Biosystems (BIOSYST), Division of Mechatronics, Biostatistics and Sensors (MeBioS), University of Leuven (KU Leuven), Kasteelpark Arenberg 30, 3001, Leuven, Belgium.}
%Tel. +32 16 377089, Fax: +32 16 321994 \\ e-mail: erkan.kayacan@biw.kuleuven.be
\address[NTU]{School of Mechanical \& Aerospace Engineering, Nanyang Technological University, 50 Nanyang Avenue, Singapore 639798. Tel. +65 6790 5585 , Fax: +65 6792 4062 \\ erdal@ntu.edu.sg}

\begin{abstract}
More efficient agricultural machinery is needed as agricultural areas become more limited and energy and labor costs increase. To increase their efficiency, trajectory tracking problem of an autonomous tractor, as an agricultural production machine, has been investigated in this study. As a widely used model-based approach, model predictive control is preferred in this paper to control the yaw dynamics of the tractor which can deal with the constraints on the states and the actuators in a system. The yaw dynamics is identified by using nonlinear least squares frequency domain system identification. The speed is controlled by a proportional-integral-derivative controller and a kinematic trajectory controller is used to calculate the desired speed and the desired yaw rate signals for the subsystems in order to minimize the tracking errors in both the longitudinal and transversal directions. The experimental results show the accuracy and the efficiency of the proposed control scheme in which the euclidean error is below $40$ cm for time-based straight line trajectories and $60$ cm for time-based curved line trajectories, respectively.
\end{abstract}


\end{frontmatter}

\section{Introduction}
One of the most important tasks in tractor operation is the accurate steering during field operations, e.g. accurate trajectory following during tillage, to avoid damaging the crop or planting when there is no crop yet. Besides, the rows must be parallel, and the distance differences between them must be equal with respect to each other during the planting. Moreover, the tractor has to cover the full field without overlap during other operations. However, the steering accuracy decreases when the operator gets tired or does more actions than driving the tractor like operating/controlling the implements. In order to automate the trajectory following problem and also increase the steering accuracy, several automatic guidance systems have been developed to avoid the problems mentioned above.

There are various reasons why the control of tractors with a high efficiency is a challenging task. First, an autonomous tractor can be configured with different types of implements and also encounter various environmental conditions (such as humidity, temperature, etc.) during field operations. In such conditions, there is always a trade-off between performance and robustness when a conventional controller, e.g. proportional-integral-derivative (PID) controller, is used. Since conventional controllers have time invariant coefficients and do not have the ability to adapt to changing conditions, they are not appropriate to be used in such agricultural production machines. Second, these machines show many nonlinear behaviors such as saturation, dead-time and time lags, which are difficult to handle with conventional control algorithms. Third, tractor navigation involves two subsystems, namely: the yaw dynamics and the longitudinal dynamics which make the control operation more challenging. There is also interaction apart from the hydraulic driveline as a change in the longitudinal speed will change the yaw dynamics and vice versa.

In model-based control, the control performance highly depends on the accuracy of the model describing the system behavior. In the last decade, several models have been proposed where the yaw dynamics of a wheeled vehicle are described with a bicycle model. Simple kinematic models have been proposed in \citep{oconnorAutomatic}. These models neglect the side-slip of the tires and the dynamics of the steering actuator. Therefore, they are not appropriate for slippery surfaces which are common in field conditions with loose or wet soil. As a solution to this problem, a bicycle dynamics model which takes the lateral forces into account is proposed in \citep{oconnorthesis}. As the effect of side slip can be taken into account by this model, it covers a range of slippery and hard surfaces. However, in the previous approach, side slip angles cannot be calculated when the longitudinal speed is equal to zero. As a solution to this problem, the relaxation length approach is proposed in \citep{karkee} to calculate the side-slip angles more accurately. Bevly \citep{Bevly} reported that the relaxation length for only the front tire is adequate in order to model the real-time system.

Modeling of side-slip angle, which is the difference between the real and effective steering angle, and determining cornering stiffness values are very important steps in analyzing the yaw dynamics of autonomous vehicles. In \citep{Fang}, the cornering stiffness is estimated by a robust adaptive Luenberger observer and a sliding mode controller is designed based-on chained system theory. The proposed controller and observer were reported to be robust to time varying lateral disturbances and also inaccurate side-slip angles. As an alternative approach to controlling the agricultural production machines, model reference adaptive control approaches have been proposed in \citep{Derrick}. It is observed that the model reference adaptive control algorithm is able to adapt itself to various implementation configurations properly to control lateral position of a tractor for a straight path. In \citep{Gartley}, the effect of the hitch point loading on the tractor dynamics is investigated by using a cascaded estimator approach. The experimental results show that the online estimation for the changes in the system provides the ability of adapting the controller gain to maintain the consistent yaw dynamic control of the tractor.

Model predictive control (MPC) has been widely used in the chemical process industry since the 1980s. The main goal of MPC approach is to minimize a performance criterion with respect to constraints of a system's inputs and outputs. The future values of the system are calculated based on a model. The main advantages of MPC over conventional controllers for the control of agricultural machines are the ability to deal with constraints and with multi-input-multi-output controllers. Several successful applications on agricultural production machines have been reported in literature. An MPC design was implemented on the cruise control of a combine harvester \citep{coen} in which the speed model was developed based on relating the engine speed and the current to the hydraulic pump to the longitudinal speed. The engine speed and the pump settings were controlled simultaneously and this approach was tested experimentally on a New Holland combine harvester. The experimental results show that a satisfactory acceleration performance can be achieved even by keeping the engine speed low. In \citep{MPC1}, an MPC strategy is described for the control of an autonomous tractor by using an extended kinematic model. This control scheme has been tested experimentally on a farm tractor whose realtime localization is achieved relying solely upon a real-time kinematic (RTK) global positioning system (GPS). However, the control accuracy is limited, because the model used is a kinematic model, and thus neglects the dynamic behaviour of the system. As an extension to MPC, a nonlinear MPC (NMPC) is proposed to obtain better lateral position accuracy of a tractor-trailer system in \citep{Backman}. The lateral position error of the trailer was reported to be less than 10 cm in straight paths for a space-based trajectory in real-time experiments. Moreover, centralized, decentralized and distributed NMPC approaches have respectively been proposed in \citep{erkanCeNMPC,erkanDiNMPC,erkanDeNMPC}. The drawback of these studies is the same as \citep{MPC1} which is that the model used does not include the dynamic behaviour of the system. Another NMPC algorithm is proposed for the yaw dynamics control of an autonomous vehicle in \citep{Canale1}. Although it was reported that the proposed controller would allow to use hard constraints for obstacle avoidance strategies, it does not include any real time experiments. Since more advanced control algorithms and mathematical models bring not only more accuracy, but also more computational burden to the real time systems, there always exists a trade-off between the complexity of the method and the computational efficiency of the overall system.

The main contributions of this study beyond the state of the art are modeling the yaw dynamics of an autonomous tractor considering various definitions of side slip angles and controlling it with good computational efficiency. In order to achieve this, first, the yaw dynamics model of the autonomous tractor has been derived, the model structures have been validated, and model parameters have been estimated by using frequency response function (FRF) measurements. Finally, the nonlinear least square (NLS) frequency domain identification (FDI) approach is used to obtain the model parameters to determine which model is better for the tractor at hand. After the identification of the yaw dynamics, an MPC controller for the yaw dynamics is designed based on the identified model. Then, this yaw dynamics controller has been combined with a kinematic controller for the trajectory tracking in which the kinematic controller is used for compensating the errors both in the x- and y-axes.

This paper is organized as follows: The experimental set-up is described in Section II. The kinematic model of the system and the mathematical model of the yaw dynamics are presented in Section III. In Section IV, the identification of the yaw dynamics is described. In Section V, the basics of the implemented MPC approach are given. The overall control structure and the real-time experimental results are presented in Section VI. Finally, some conclusions are drawn from this study in Section VII.

%%%%%%%%%%%%%%%%%%%%%%%%%%%%%%%%%%%%%%%%%%%%%%%%%%%%%%%%%%%%%%%%%%%%%%%%%%%%%%%%%%%%%%%%%%%%%%%%%%%%%%%%%%%%%%%%%%%%%%%%%%%%%

\section{Experimental Set-up Description}
\begin{figure}[t!]
\centering
  \includegraphics[width=3in]{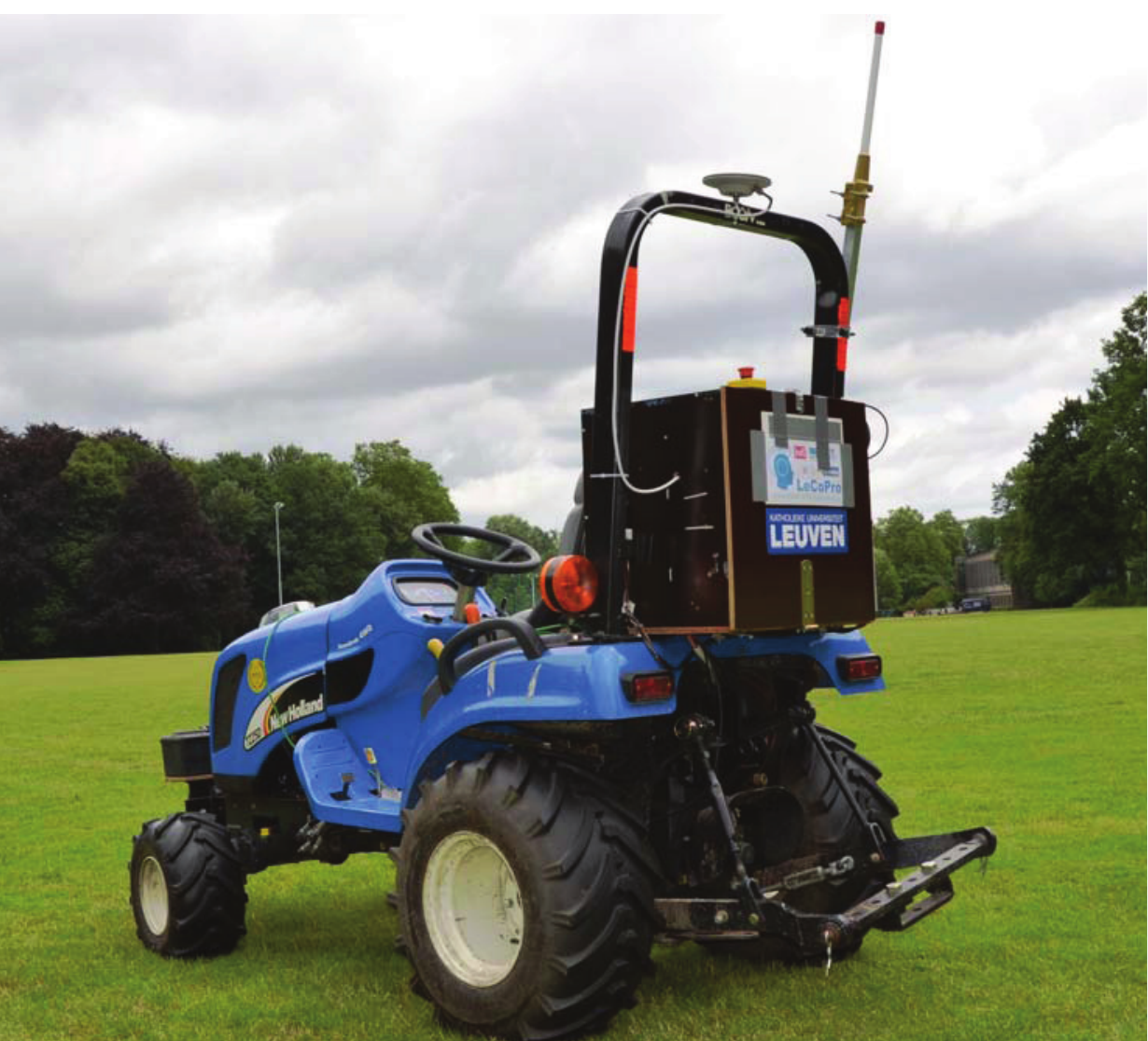}\\
  \caption{The experimental set-up (CNH TZ25DA)}
  \label{tractor1}
\end{figure}

The aim of this study is to track a time-based trajectory with a small agricultural tractor shown in Fig. \ref{tractor1}. The GPS antenna is located straight up the center of the tractor rear axle to provide highly accurate position information for the autonomous tractor. The height of the antenna is 2 m above ground level. It is connected to a Septentrio AsteRx2eH RTK-DGPS receiver (Septentrio Satellite Navigation NV, Leuven, Belgium) with a specified position accuracy of 2 cm at a 20-Hz sampling frequency. The Flepos network supplies the RTK correction signals via internet by using a \emph{Digi Connect WAN 3G} modem.

\begin{figure}[t!]
\centering
  \includegraphics[width=3.4in]{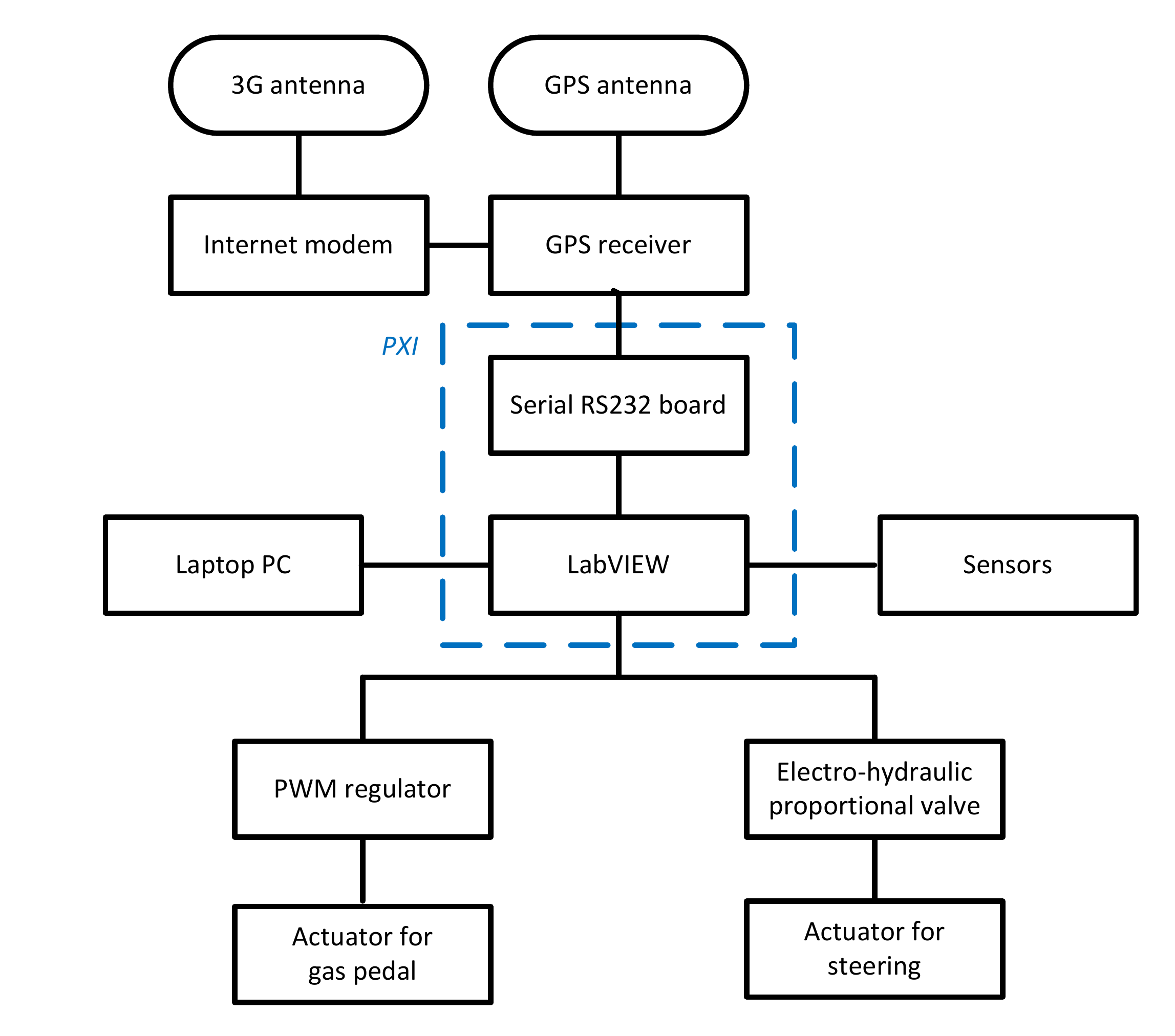}\\
  \caption{Block diagram of hardware}
  \label{blockdiagramhardware2}
\end{figure}

The block diagram of the hardware is shown in Fig. \ref{blockdiagramhardware2}. The GPS receiver and the internet modem are connected to a real time operating system (PXI platform, National Instruments Corporation, Austin, TX, USA) through an RS232 serial communication. The PXI system acquires the steering angle, the GPS data and controls the tractor by sending messages to actuators. A laptop connected to the PXI system by WiFi functions as the user interface of the autonomous tractor. The algorithms are implemented in $LabVIEW^{TM}$ version 2011 (National Instruments, Austin, TX, USA). They are executed in real time on the PXI and updated at a rate of 20-Hz.

The designed MPC in Section \ref{sectionMPC} calculates the desired steering angle for the front wheels, and a low level controller, a PI controller in our case, is used to control the steering mechanism. In the inner closed loop, the steering mechanism is controlled by using an electro-hydraulic valve with a maximal flow of 12 liter/min. The electro-hydraulic valve characteristics are highly nonlinear and include a saturation and a dead-band region. The voltage limited between $0-12$ volt and the steering angle limited between $\pm45^{\circ}$ became the input and the output for the steering system, respectively. The angle of the front wheels is measured using a potentiometer mounted on the front axle, yielding an angle measurement resolution of $1^{\circ}$.

The speed of the tractor is controlled by using an electro-mechanic valve. There are two PID type controllers in speed control. The PID controllers in the outer closed-loop and the inner closed-loop are generating the desired pedal position with respect to the speed of the tractor and voltage for the electro-mechanic valve with respect to the pedal position, respectively. Since the measured speed coming from the GPS is noisy, a discrete Kalman Filter (KF) was used to reduce noise. A position-velocity model described in \citep{Brown} was used where vehicle velocity is assumed as a random-walk process. The KF assumes that the vehicle moves with a constant velocity between discrete-time steps. The state vector of the model used in the KF and the state transition matrix are written as follows:

\begin{eqnarray}
\label{kinematicmodelKF}
\widehat{\textbf{x}}_{k+1} & = & \Phi(T_{s}) \widehat{\textbf{x}}_{k} \nonumber \\
  & = & \left[
  \begin{array}{cccc}
   1 & T_{s} & 0 & 0 \\
   0 & 1 & 0 & 0 \\
   0 & 0 & 1 & T_{s}  \\
   0 & 0 & 0 & 1  \\
  \end{array}
  \right]
   \left[
  \begin{array}{c}
   x_{k}  \\
   v_{x,k}  \\
   y_{k}  \\
   v_{y,k}  \\
  \end{array}
  \right]
\end{eqnarray}
where $\Phi(T_{s})$, $v_{x,k}$ and $v_{y,k}$ are the state transition matrix and velocities coming from the GPS, respectively.

%%%%%%%%%%%%%%%%%%%%%%%%%%%%%%%%%%%%%%%%%%%%%%%%%%%%%%%%%%%%%%%%%%%%%%%%%%%%%%%%%%%%%%%%%%%%%%%%%%%%%%%%%%%%%%%%%%%%%%%%%%

\section{Mathematical Model of an Autonomous Tractor}

\subsection{Kinematic Model}

\begin{figure}[h!]
\centering
  \includegraphics[width=3in]{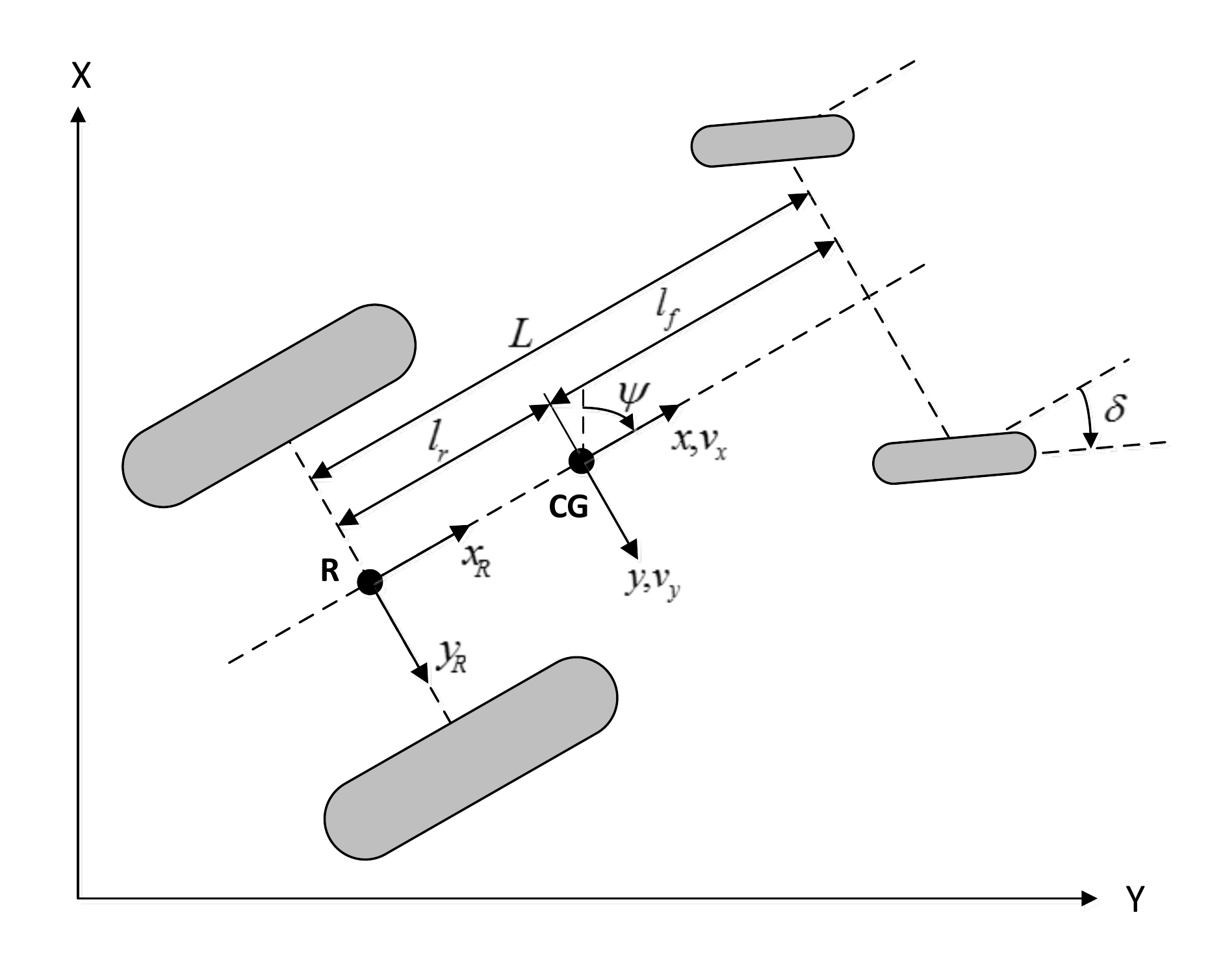}\\
  \caption{Kinematic model for an autonomous tractor}
  \label{kinematic}
\end{figure}

The schematic diagram of an autonomous tractor is illustrated in Fig. \ref{kinematic}. The linear velocities $\dot{x}$ and $\dot{y}$ at the rear axle of the tractor (point R) are written as follows:

\begin{eqnarray}\label{kinematicmodeltractor}
\dot{x}_{R} & = & v_{x} \cos{\psi} \nonumber \\
\dot{y}_{R} & = &  v_{x} \sin{\psi} \nonumber \\
\dot{\psi}_{R}  & = & \frac{v_{x} \tan{\delta}}{L}
\end{eqnarray}
where $v_{x}$, $\psi$, $\delta$ and $L$ represent the longitudinal velocity, the yaw angle of the tractor, the steering angle of the front wheel, the distance between the front axle and the rear axle of the tractor, respectively.

Instead of considering the point R in the dynamic equations and the trajectory tracking control, the center of gravity (CG) is preferred. Thus, the linear velocities $\dot{x}$ and $\dot{y}$ are projected onto the CG as follows:

\begin{eqnarray}\label{kinematicmodeltractor2}
\dot{x} & = & v_{x} \cos{\psi} - v_{y} \sin{\psi} \nonumber \\
\dot{y} & = & v_{x} \sin{\psi} + v_{y} \cos{\psi}
\end{eqnarray}
where $v_{y}$ is the lateral velocity of the tractor at the CG.

\subsection{Modeling of the Yaw Dynamics}\label{sectionmodeling}

\begin{figure}[b!]
\begin{center}
  \includegraphics[width=3.2 in]{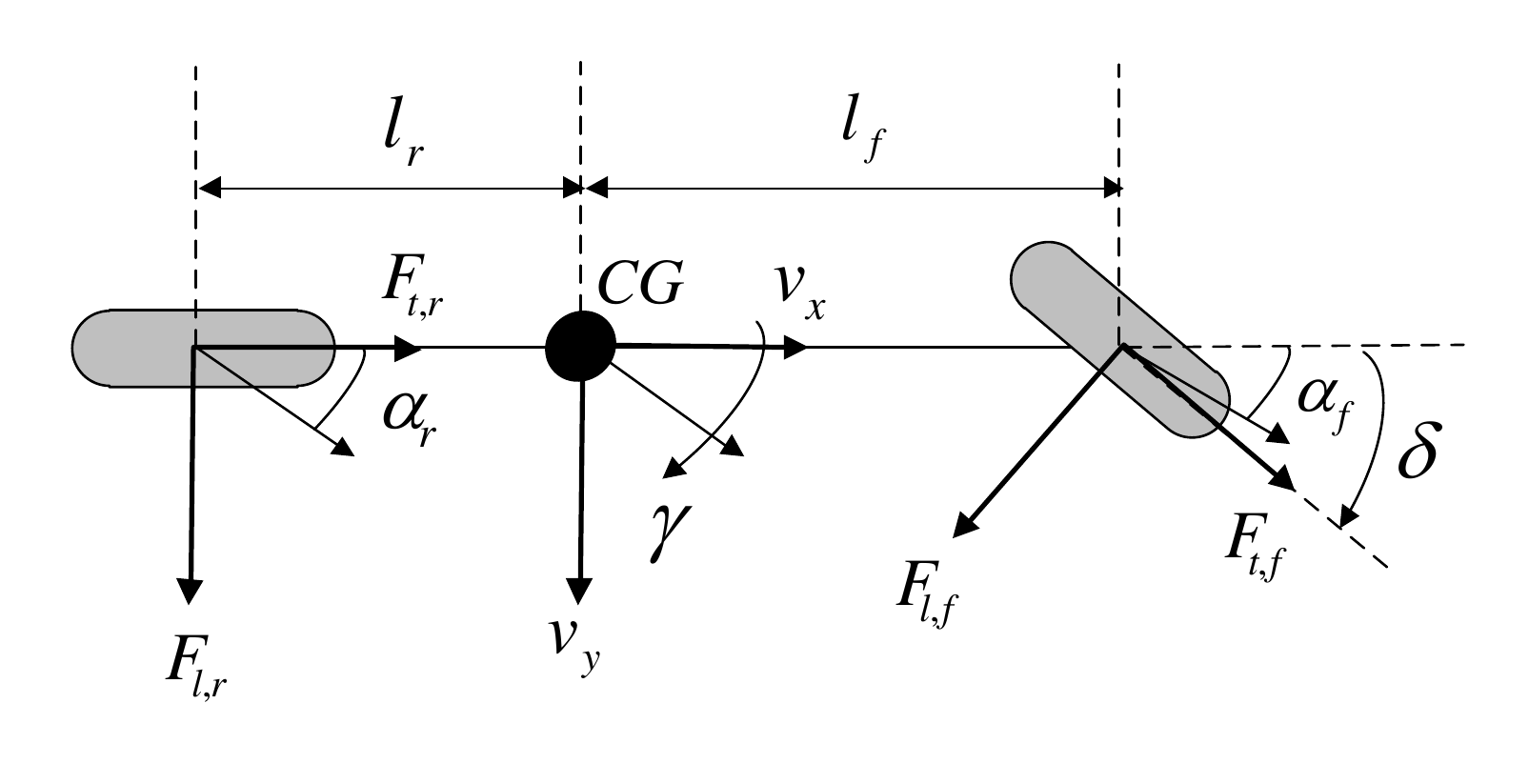}\\
  \caption{Bicycle dynamics model for a tractor system: velocities, side slip angles and forces on the rigid body of the system }\label{tractorschematic}
   \end{center}
\end{figure}

As the driving speed of the tractor is rather limited, it is reasonable to assume that the lateral forces on the right and left wheels are equal to each other and can be summed. Therefore, the tractor is modelled in 2D as a bicycle system. The velocities, the sideslip angles and the forces on the rigid body of an autonomous tractor are schematically illustrated in Fig. \ref{tractorschematic}. The yaw dynamics models are derived based on the following assumptions:

\begin{itemize}
  \item The traction forces are neglected,
  \item The aerodynamic forces are neglected,
  \item The tire moments are small, such that these can be neglected,
  \item The pitch and roll dynamics are neglected.
\end{itemize}

The notations used in the following (see also Fig. \ref{tractorschematic}) are summarized in Table \ref{nomenclature}.

\begin{table}[h!]
\caption{NOMENCLATURE}
\centering
\begin{tabular}{c l}
\hline
$m$ & Mass\\
$I$ & Moment of inertia of the tractor around the vertical axis\\
$v_{x}$ & Longitudinal velocity of the CG \\
$v_{y}$ & Lateral velocity of the CG\\
$\psi$ & Yaw angle of the tractor \\
$\gamma$ & Yaw rate of the tractor \\
$\delta$ & Steering angle of the front wheels \\
$l_{f}$ & Distance between the front axle and the CG \\
$l_{r}$ & Distance between the rear axle and the CG \\
$L$ & Distance between the front axle and the rear axle \\
$F_{t,f}$ & Traction force on the front wheels\\
$F_{t,r}$ & Traction force on the rear wheels\\
$F_{l,f}$ & Lateral force on the front wheels\\
$F_{l,r}$ & Lateral force on the the rear wheels\\
$C_{\alpha, f}$ & Cornering stiffness of the front wheels\\
$C_{\alpha, r}$ & Cornering stiffness of the rear wheels\\
$\alpha _{f}$ & Side slip angle of the front  wheels\\
$\alpha_{r}$ & Side slip angle of the rear wheels\\
$\sigma _{f}$ & Relaxation length of the front wheels\\
$\sigma _{r}$ & Relaxation length of the rear wheels\\
\hline
\end{tabular}
\label{nomenclature}
\end{table}

\subsubsection{Vehicle Dynamics}
%
%The longitudinal equation of motion of the tractor can be written as follows:
%\begin{equation}\label{longitudinalmotionoftractor}
%   m (\dot{v}_{x} - v_{y} \gamma ) = F_{t,f} \cos\delta - F_{l,f} \sin\delta + F_{t,r}
%\end{equation}
%where $\gamma$, $m$, $F_{t,f}$, $F_{l,f}$, $F_{t,r}$ represent the yaw rate and the mass of the tractor, the traction and lateral forces on the front wheel, the traction force on the rear wheel, respectively.

The lateral motion of the tractor is written as follows:

\begin{equation}\label{lateralmotionoftractor}
   m (\dot{v_{y}} + v_{x} \gamma ) = F_{t,f} \sin\delta + F_{l,f} \cos\delta + F_{l,r}
\end{equation}
where $\gamma$, $m$, $F_{t,f}$, $F_{l,f}$, $F_{l,r}$  represent the yaw rate and the mass of the tractor, the traction and lateral forces on the front wheel, and the lateral force on the rear wheel.

The yaw motion of the tractor is written as follows:

\begin{eqnarray}\label{yawmotionoftractor}
   I \dot{\gamma} & =  & l_{f} ( F_{t,f} \sin\delta + F_{l,f} \cos\delta ) - l_{r} F_{l,r}
\end{eqnarray}
where $l_{f}$, $l_{r}$ and $I$ represent the distance between the front axle and the CG of the tractor, and the distance between the rear axle and the CG of the tractor, the moment of inertia of the tractor. Since it is a time consuming process to calculate the inertial moment of the tractor, an approximate value for the inertial moment can be calculated as follows \citep{Garrott}:

\begin{equation}\label{inertiamoment}
 I= m l_f l_r
\end{equation}

\subsubsection{Tire Model}

The lateral tire forces are calculated in a linear model in which they are assumed to be proportional to the slip angles \citep{Piyabongkarn,Geng} :

\begin{equation}\label{lateralforces}
    F_{l,i} = - C_{\alpha,i}  \alpha_{i}  \quad \quad i=\{f,r\}
\end{equation}
where $C_{\alpha,i}$ and $\alpha_{i}$ $i=\{f,r\}$, represent the cornering stiffnesses of the tires and the side-slip angles, respectively.

%This linear model is valid if the side-slip angles are smaller than 10 degress. Since a small side-slip angles assumption has been made in Section \ref{sectionmodeling}, this linear model is a proper candidate.

The tire side-slip angles must be calculated in order to determine the slip forces. The side slip angles of the front and rear tires have been considered in a linear form, and they are written as follows:

\begin{eqnarray}\label{eq9}
   \alpha_{f}  & = & \frac{v_{y}+  l_f \gamma}{v_{x}} - \delta \label{sideslipanglef}  \\
   \alpha_{r}  & = & \frac{v_{y} -  l_r \gamma}{v_{x}} \label{sideslipangler}
\end{eqnarray}
As can be seen from the equations above, the side slip angles cannot be calculated when the longitudinal speed is zero. As a solution to this problem, the relaxation length is defined as the amount a tire rolls to reach the steady state side slip angle. Previous research in vehicles suggests that the relaxation length of a tire plays a very important role in the steering motion at high velocities \citep{Owen,Crolla}. This result motivates us to use this approach for the agricultural vehicles due to the fact that the calculation of the side-slip angles in (\ref{sideslipanglef}) and (\ref{sideslipangler}) for very low velocities can go to infinity. Since a tire generates the steady state side slip angle simultaneously, a first order mathematical model is used to describe the slip angle dynamics through the relaxation length. A first order differential equation for the side slip angle can be written as follows:

\begin{equation}\label{eq10}
\dot{\alpha} = \frac{v_{x}}{\sigma} (\alpha_{0} - \alpha)
\end{equation}

A relaxation length of 1.5 times the tire radius has been proposed for agricultural vehicles as it allows to obtain similar changes for a similar increase in velocity \citep{Bevly}. For passenger vehicles which have higher velocity than agricultural vehicles, a factor larger than 2 is typically selected \citep{Loeb}.

By combining equations (\ref{sideslipanglef}), (\ref{sideslipangler}) and (\ref{eq10}), the time derivatives of the side slip angles of the front and rear wheels can be written as follows:

\begin{eqnarray}
   \dot{\alpha}_{f} & = & \frac{v_{y}+  l_f \gamma - v_{x} (\delta  + \alpha_{f})}{\sigma_{f}}  \label{sideslipangleswrlf} \\
   \dot{\alpha}_{r} & = & \frac{v_{y} -  l_r \gamma - v_{x} \alpha_{r}}{\sigma_{r}} \label{sideslipangleswrlr}
\end{eqnarray}
where $\sigma_{f}$ and $\sigma_{r}$ represent the relaxation length of the front and rear tires of the tractor, respectively.

\subsubsection{Equations of Yaw Motion}\label{sectioneqyaw}

The tractor is driven on the field with a constant longitudinal velocity as generally required in automatic guidance of agricultural vehicles. For this reason, the time derivative of the longitudinal velocity $\dot{v}_{x}$ can be set to zero and the longitudinal velocity $v_{x}$ can be assumed as a parameter. It is also assumed that the steering angle is sufficiently small to justify linearization of the equations. Thus, (\ref{lateralmotionoftractor}) and (\ref{yawmotionoftractor}) can be written considering (\ref{lateralforces}) as follows:

\begin{eqnarray}\label{linearequationsofyawmotion}
   m \dot{v}_{y} & = & - m v_{x} \gamma - C_{\alpha,f}  \alpha_{f} - C_{\alpha,r}  \alpha_{r}  \nonumber \\
   I \dot{\gamma} & = & - l_{f} C_{\alpha,f}  \alpha_{f} + l_{r} C_{\alpha,r}  \alpha_{r}
\end{eqnarray}

By using different considerations, three different transfer functions can be written. Firstly, the traditional bicycle model for the yaw motion of the autonomous tractor can be obtained by combining (\ref{sideslipanglef}), (\ref{sideslipangler}) and (\ref{linearequationsofyawmotion}). The relation between the yaw rate of the autonomous tractor and the steering angle of the front wheels can be written in transfer function form as follows:

\begin{equation}\label{TBMtf}
G_{TB}(s)=\frac{b^{\star}_{1} s + b^{\star}_{0}}{a^{\star}_{2} s^{2} + a^{\star}_{1} s + a^{\star}_{0}}
\end{equation}

Secondly, the bicycle model with relaxation length approach for only the front wheel can be obtained by combining (\ref{sideslipangler}), (\ref{sideslipangleswrlf}) and (\ref{linearequationsofyawmotion}). The relation between the yaw rate of the autonomous tractor and the steering angle of the front wheels can be written in transfer function form as follows:

\begin{equation}\label{BMwrltff}
G_{RLF}(s)=\frac{b^{\diamond}_{1} s + b^{\diamond}_{0}}{a^{\diamond}_{3} s^{3} + a^{\diamond}_{2} s^{2} + a^{\diamond}_{1} s + a^{\diamond}_{0}}
\end{equation}

Thirdly, the bicycle model with the relaxation length approach for the front and the rear wheels can be obtained by combining (\ref{sideslipangleswrlf}), (\ref{sideslipangleswrlr}) and (\ref{linearequationsofyawmotion}). The relation between the yaw rate of the autonomous tractor and the steering angle of the front wheels can be written in transfer function form as follows:

\begin{equation}\label{BMwrltffr}
G_{RLFR}(s)=\frac{b^{\ast} _{2} s^2 + b^{\ast} _{1} s + b^{\ast} _{0}}{a^{\ast} _{4} s^{4} + a^{\ast} _{3} s^{3} + a^{\ast} _{2} s^{2} + a^{\ast} _{1} s + a^{\ast}_{0}}
\end{equation}

The relation between the physical parameters in the previous equations and the transfer function parameters in (\ref{BMwrltff}) and (\ref{BMwrltffr}) is shown in Appendix A. These different models might be proper for different real-time systems or different cases of the same real-time system. In Section IV these transfer functions (\ref{TBMtf}), (\ref{BMwrltff}) and (\ref{BMwrltffr}) will be fit to the empirical transfer function estimate obtained from the frequency domain experiments to decide which is most appropriate for the real-time system considered in this study.

%%%%%%%%%%%%%%%%%%%%%%%%%%%%%%%%%%%%%%%%%%%%%%%%%%%%%%%%%%%%%%%%%%%%%%%%%%%%%%%%%%%%%%%%%%%%%%%%%%%%%%%%%%%%%%%%%%%%%%%%%%

\section{Identification of the Yaw Dynamics}\label{sectionident}
\begin{figure}[h!]
\begin{center}
  \includegraphics[width=5 in]{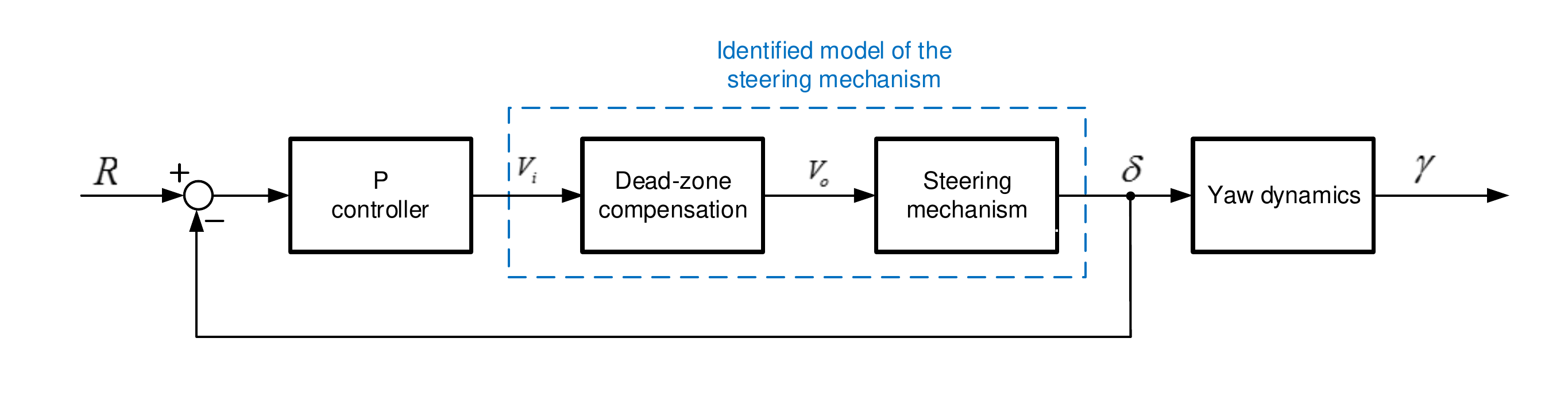}\\
  \caption{The schematic diagram of the identification processes}\label{identblockdiagram}
   \end{center}
\end{figure}

It was observed during the identification experiments that the front wheels reached their limits due to drift when the excitation signal was applied to the steering mechanism in an open-loop fashion. As a solution to this drifting problem, the system was controlled with a P controller, and then the closed-loop system was identified. The schematic diagram of the identification processes is shown in Fig. \ref{identblockdiagram}. The steering mechanism of the tractor was identified in \citep{erkanmodelleme,Kayacan2013ASCC}.

In this section, the nonlinearity of the yaw dynamics is checked by using an odd-odd multisine signal and a frequency content in which the linear contributions are dominant is determined for the identification process \citep{Schoukens,Schoukens2}. A multisine signal is applied to the steering angle controller, and the linear models with respect to Section \ref{sectioneqyaw} are obtained by using the NLS FDI method \citep{Kayacan2013ASCC}. After the frequency domain analysis, a qualitative comparison of the linear models in time-domain is given, and the effect of the longitudinal speed is shown.

\begin{figure}[h!]
\begin{center}
  \includegraphics[width=3.4 in]{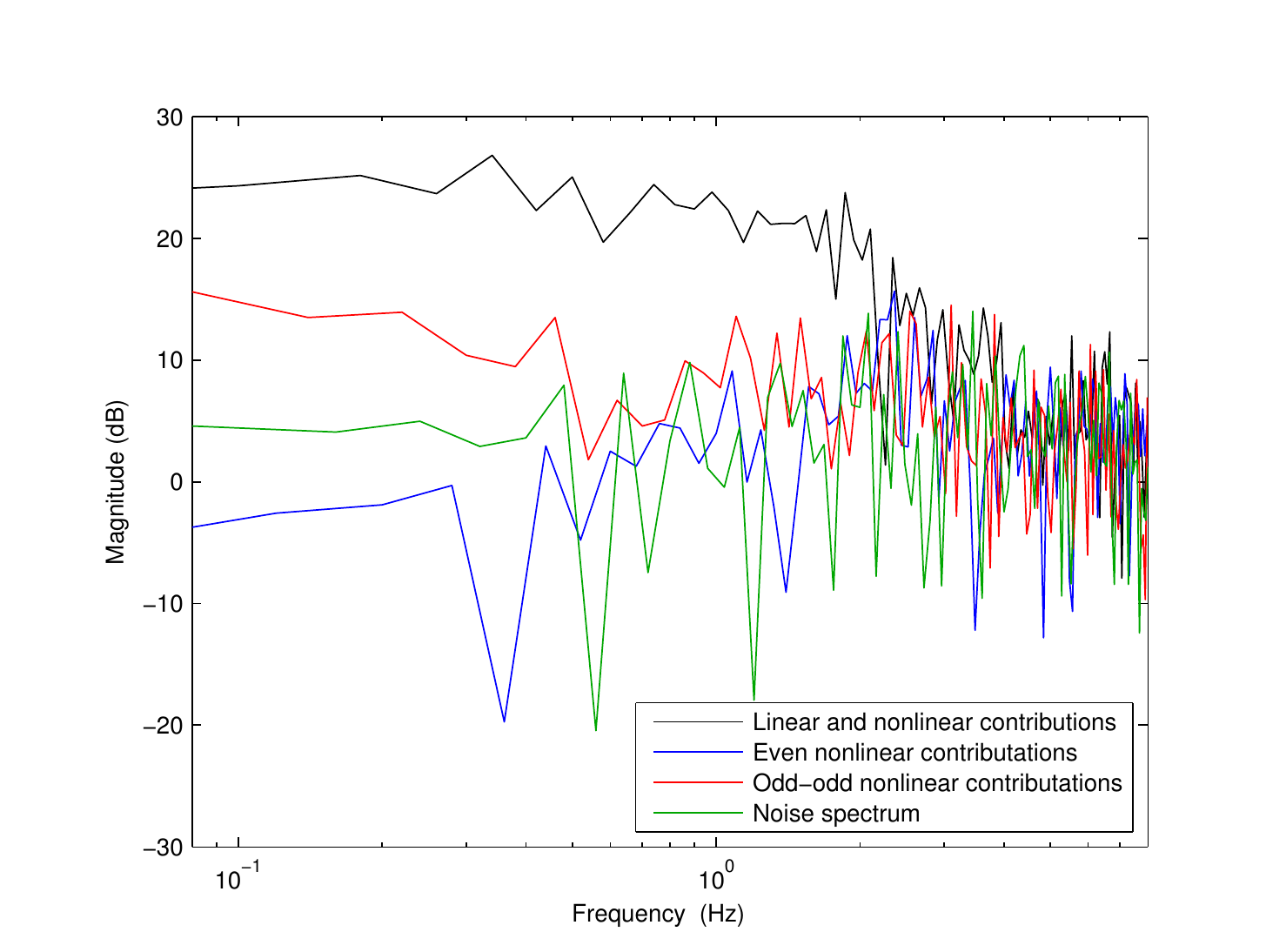}\\
  \caption{The analysis of nonlinear contributions} \label{responseoddodd_yawrate}
   \end{center}
\end{figure}

\begin{figure}[h!]
\begin{center}
  \includegraphics[width=3.4 in]{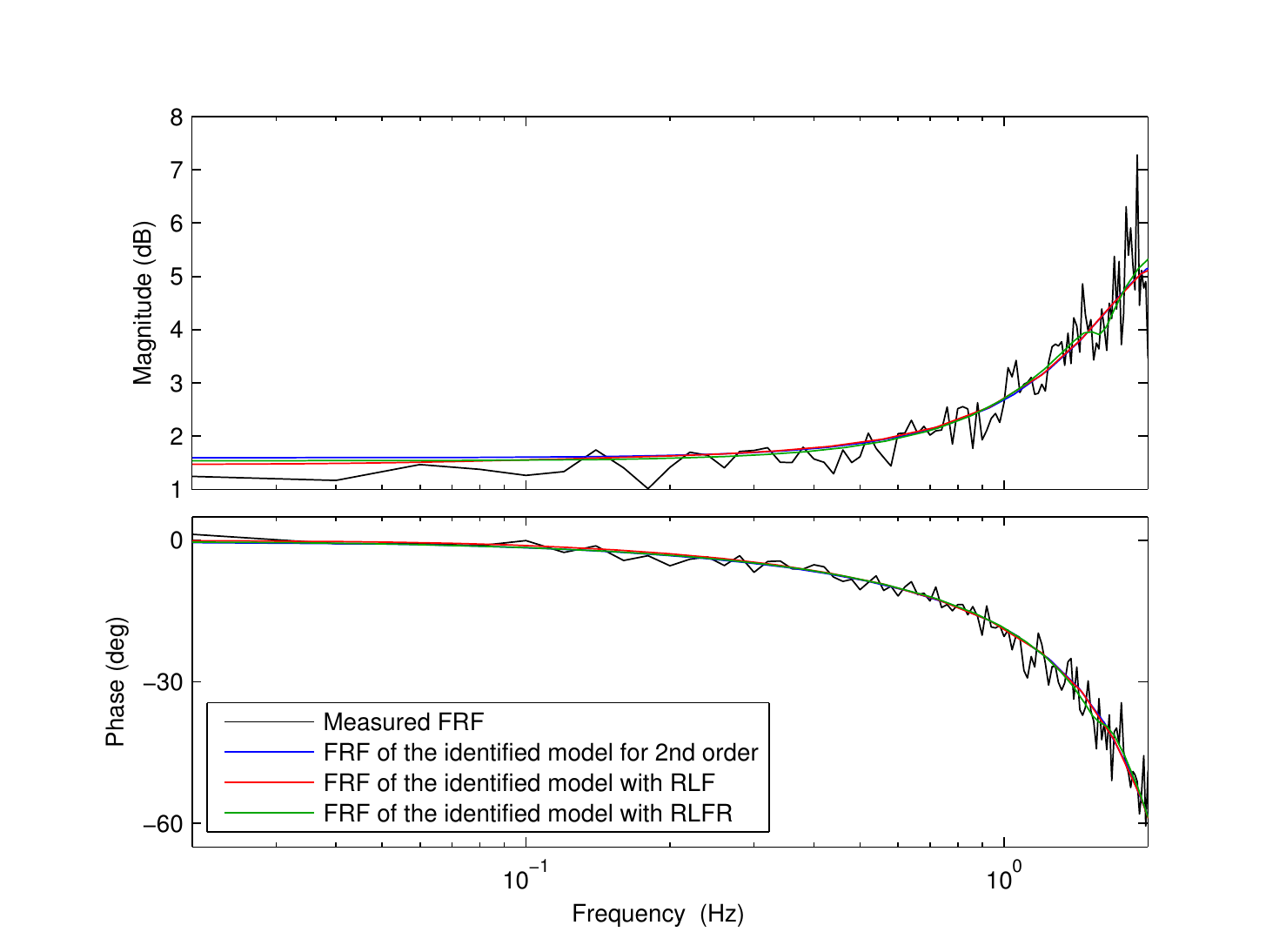}\\
  \caption{Measured FRF and FRF of the identified models}\label{NLS_yawrate}
   \end{center}
\end{figure}

After the design and the implementation of the steering angle controller, the frequency spectrum of the response of the yaw rate to a random odd-odd multisine excitation is shown in Fig. \ref{responseoddodd_yawrate}. It can be seen that the contribution of the nonlinearities to the total response is as large as the linear contribution after 2 Hz. Moreover, it is known that the noise in FRF measurements above 1.5 Hz will be due to the lack of input signal. As can be seen from Fig. \ref{NLS_yawrate}, the range till 2 Hz is not enough to capture the peak, and the magnitude is still quite high. Since the nonlinear contributions are dominant after 2 Hz, a linear model can be derived until 2 Hz.

Based-on the considerations above, a multisine signal with a frequency content between 0.02Hz and 2Hz has been applied to the system as an excitation signal. The model parameters are identified by using the NLS FDI approach based on FRF measurements. In Fig. \ref{NLS_yawrate}, the measured FRF and the FRFs of identified models are shown. As can be seen from Fig. \ref{NLS_yawrate}, since the traditional bicycle model in (\ref{TBMtf}) consists of one zero and two poles such a system cannot have a frequency response like the one in Fig. \ref{NLS_yawrate}, and thus it is not a candidate for the parameter estimation. The other two models in (\ref{BMwrltff}) and (\ref{BMwrltffr}) can be proper candidates for the parameter estimation.

After the identification process, the transfer functions of the identified yaw dynamics model are calculated respectively for the bicycle model with relaxation length approach for only the front wheel, the bicycle model with relaxation length approach for the front and the rear wheels, and a second order system as follows:

\begin{equation}\label{yawmodelRLF}
G_{RLF}(s)= \frac{292s + 177}{s^{3} + 11.6 s^2 + 249 s + 150}
\end{equation}
\begin{equation}\label{yawmodelRLFR}
G_{RLFR}(s)= \frac{279s^2 + 335s + 27860}{s^4 + 11.5 s^3 + 347s^2 + 1311 s + 23340}
\end{equation}
\begin{equation}\label{yawmodel2nd}
G_{2nd}(s)= \frac{291}{s^{2} + 10.9 s + 242}
\end{equation}

Even if the models derived in (\ref{BMwrltff}) and in (\ref{BMwrltffr}) fit with the real-time FRF measurements, these models are higher order transfer functions. As a simplified model, a second order transfer function in (\ref{yawmodel2nd}) is fitted to FRF measurements. As can be seen from Fig. \ref{NLS_yawrate}, the frequency domain response of the proposed transfer function in (\ref{yawmodel2nd}) is similar to the derived models in (\ref{yawmodelRLF}) and (\ref{yawmodelRLFR}). It can be concluded that a second order transfer function can also be used to model the yaw dynamics of the tractor at hand. It should not go unnoticed that we do not throw the physical models away by proposing the second order empirical transfer function. The reason is that the empirical model might not be suitable for different soil conditions due to the higher values for the slip. For a better understanding, (\ref{sideslipangleswrlf}), (\ref{sideslipangleswrlr}) and (\ref{linearequationsofyawmotion}) should be considered carefully in which there are four ordinary differential equations that describe the yaw dynamics behaviour of the tractor. In case of having small side-slip angles, the equations for the side-slips can be neglected. As a result, there will only be two ordinary differential equations for the yaw dynamics resulting in a second order transfer function. Another disadvantage of using the empirical model is that the side-slip angles cannot be estimated.

A qualitative comparison of the three identified models in time-domain is given in Table \ref{RMSE}. A multisine signal between 0 and 8 Hz is given to the system to check whether the models are appropriate for the real-time system at hand. As can be seen from Table \ref{RMSE}, although the parameter estimation of the three models is done until 2Hz in frequency domain, the models still give reasonable results at higher frequencies. Moreover, it can be concluded that the addition of zeros and poles do not increase the accuracy of the identified models when the side-slips are negligible.

\begin{table}[h]
\caption{Root Mean Square Error of the Identified Models}
\centering
\begin{tabular}{l c c rrrrrrr}
\hline\hline
Model & Multisine Range & RMSE
\\ [0.5ex]
\hline % inserts single-line
% Entering 1st row
& 0-2Hz & 0.0787 \\[-1ex]
\raisebox{1.5ex}{$G_{2nd}(s)$}
& 0-8Hz & 0.0854 \\[1ex]
% Entering 2nd row
& 0-2Hz & 0.0787 \\[-1ex]
\raisebox{1.5ex}{$G_{RLF}(s)$}
& 0-8Hz & 0.0854 \\[1ex]
% Entering 3rd row
& 0-2Hz & 0.0793\\[-1ex]
\raisebox{1.5ex}{$G_{RLFR}(s)$}
& 0-8Hz & 0.0860\\[1ex]
% [1ex] adds vertical space
\hline % inserts single-line
\end{tabular}
\label{RMSE}
\end{table}

The selected and estimated parameters for the bicycle model with the relaxation length approach for the front and the rear wheels are given in Table \ref{estimatedparameters}. It is assumed that the values of the mass $m$, the distance between the front axle and the CG $l_{f}$, and the distance between the rear axle and the CG $l_{r}$ are known. The inertia moment $I$ is calculated based-on (\ref{inertiamoment}). The values for the cornering stiffness and the relaxation length of the front and the rear wheels are approximately estimated based-on (\ref{yawmodelRLFR}). On the other hand, since it is observed that the parameters are not realistic for the bicycle model with the relaxation length approach for the front wheel, they have not been able to be estimated. During the parameter estimation process, it has been observed that there are more than one solution for a specific parameter. When two separate solutions resulted in roughly the same values, the estimated parameters values have been said to be realistic.

\begin{table}[h]
\caption{Numerical Values of the Parameters}
\centering
\begin{tabular}{l l l}
\hline\hline
Parameter & Unit & Value
\\ [0.5ex]
\hline
$m$ & $kg$ & 700 \\
$I$ & $kg$ $m^2$  & 280 \\
$l_{f}$ & $m$ & 1 \\
$l_{r}$ & $m$ & 0.4 \\
$C_{\alpha, f} $ & $N/rad$ & $8000\pm500$ \\
$C_{\alpha, r} $ & $N/rad$ & $90000\pm7000$ \\
$\sigma _{f} $ & $m$ & $ 0.1942$ \\
$\sigma _{r} $ & $m$ & $ 1.6657$ \\
\hline
\end{tabular}
\label{estimatedparameters}
\end{table}

\subsection{The effect of longitudinal velocity}\label{subsectioneffectllv}
\begin{figure}[htb]
\centering
\subfigure[ ]{
\includegraphics[width=1.6in]{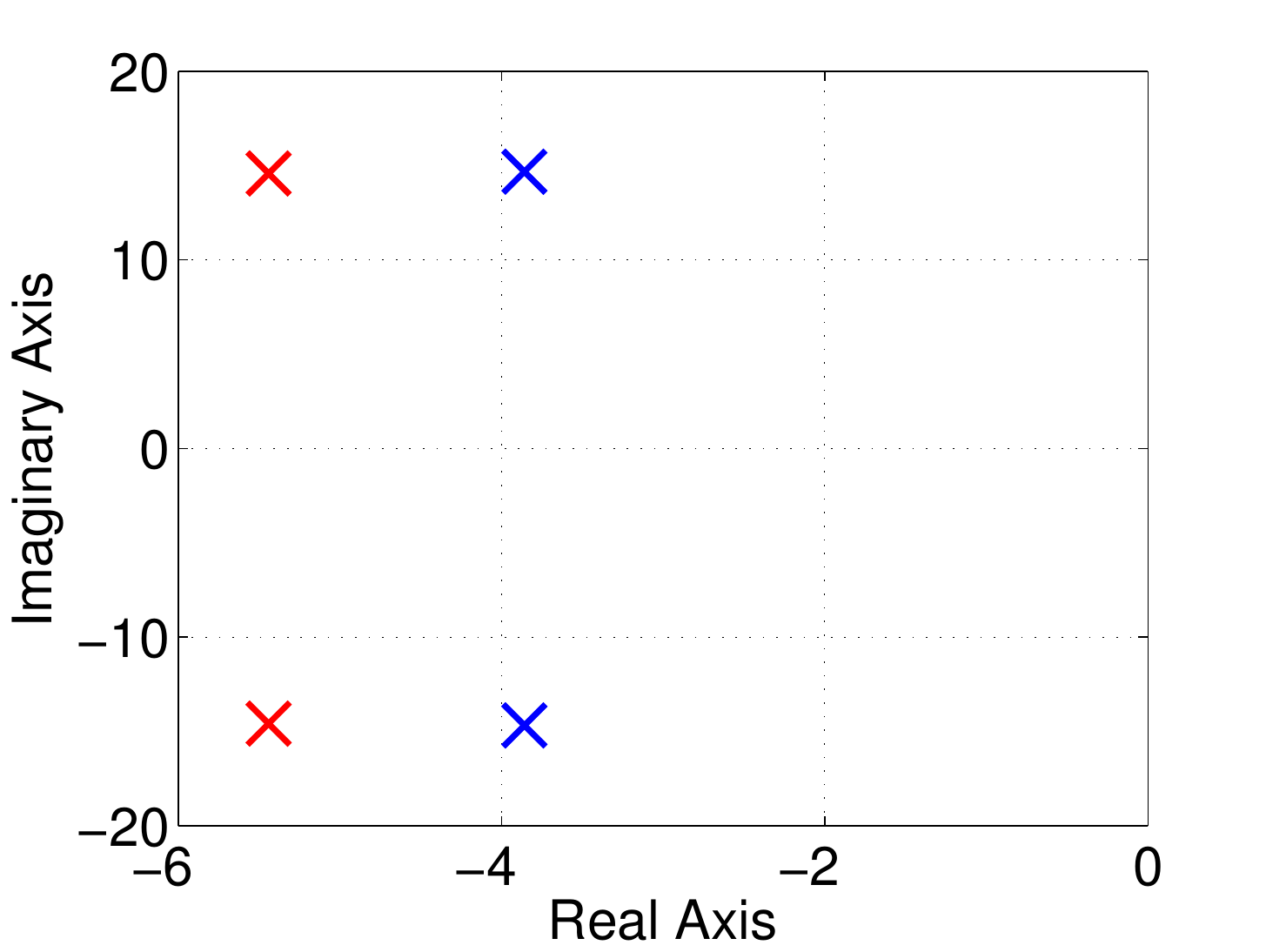}
\label{}
}
\subfigure[ ]{
\includegraphics[width=1.6in]{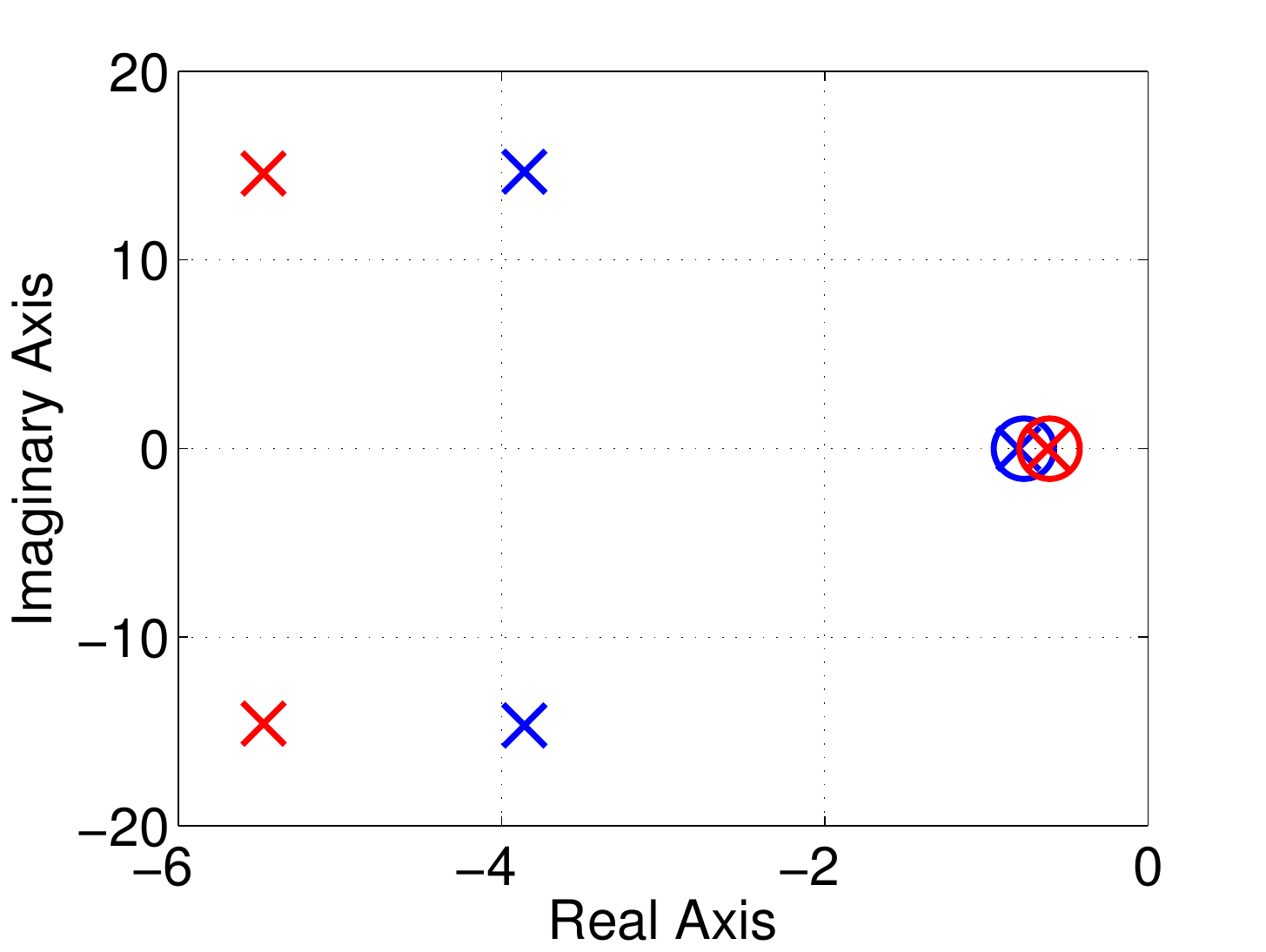}
\label{}
}
\label{}
\subfigure[ ]{
\includegraphics[width=1.6in]{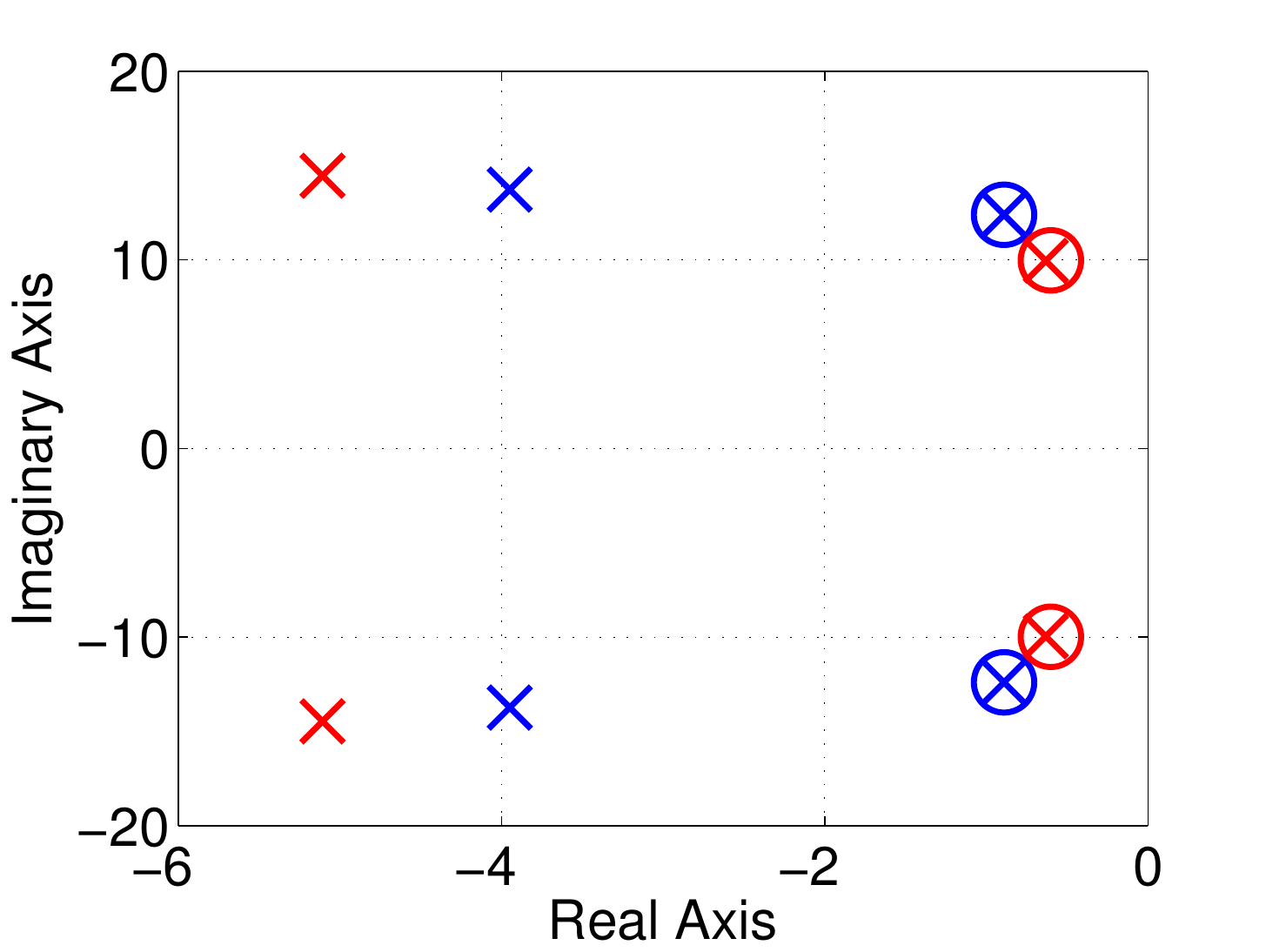}
\label{}
}
\caption[Optional caption for list of figures]{Poles and zeroes at 1 m/s (blue) and 2 m/s (red): (a) 2nd order model (b) the bicycle model with RLF (c) the bicycle model with RLFR} \label{poleszeroes}
\end{figure}

In order to be able to analyze the effect of the linear longitudinal velocity on the identified linear models, an additional experiment has been performed in which the linear velocity is varied from $1$ m/s to $2$ m/s. In Figure \ref{poleszeroes}, it is illustrated how the poles and the zeroes of the three models mentioned above change with respect to the longitudinal velocity. As can be seen from Fig. \ref{poleszeroes}, the bicycle model with RLFR and the bicycle model with RLF have pole-zero cancelation, and the poles go to the left on the s-plane, which is an expected case from the given transfer functions in (\ref{BMwrltff}) and (\ref{BMwrltffr}), when the longitudinal velocity increases.

%%%%%%%%%%%%%%%%%%%%%%%%%%%%%%%%%%%%%%%%%%%%%%%%%%%%%%%%%%%%%%%%%%%%%%%%%%%%%%%%%%%%%%%%%%%%%%%%%%%%%%%%%%%%%%%%%%%%%%%%%%

\section{Model Predictive Control}\label{sectionMPC}

MPC controllers predict the future system behavior based-on the system model, and calculate the optimal input sequence based on these predictions \citep{Maciejowski}. As it was stated in the previous section, among the identified models, the second-order empirical model gives similar performance accuracy with the other mentioned physical models to represent the autonomous tractor at hand. Thus, the second-order empirical model is used for the MPC design in this study. The objective function consists of a function of the states, the outputs and the inputs of the system. The control action is calculated by minimizing the cost function subject to the predicted behavior of the model and the system constraints. The states and the outputs are predicted over a given prediction horizon. The main equality constraint is the system model, and the inequality constraints are the state constraints, the output and the input constraints (actuators limits). In our case, there are no constraints on the system states, but the constraint on the input to the system, which is the steering angle, is defined in (\ref{constraints}).

A discrete-time linear invariant state-space model can be written as follows:

\begin{eqnarray}\label{statespacempc}
x(k+1) & = & A x(k) + B u(k) \nonumber \\
y(k) & = & C x(k) + D u(k)
\end{eqnarray}
where $x(k) \in \mathbb{R}^{n}$, $y(k) \in \mathbb{R}^{p}$ and $u(k) \in \mathbb{R}^{m}$ are the state, output and input variables, respectively. The pre-known matrices $A$, $B$, $C$ and $D$ are calculated considering the sampling time of the real-time system by using (\ref{yawmodel2nd}).

The constraints are written for all $k \geq 0$ as follows:

\begin{eqnarray}\label{constraints}
-45 \; degrees  \leq & u(k) & \leq 45  \;degrees \nonumber \\
-55 \; degrees/s  \leq & \Delta u(k) & \leq  55\; degrees/s
\end{eqnarray}

The cost function in its general form is written as follows:

\begin{equation}\label{costfunction}
J\big(U, x(k)\big)=\displaystyle\sum\limits_{i=0}^{N_{p}} x^{T} _{k+i|k} Q x_{k+i|k} + \displaystyle\sum\limits_{i=0}^{N_{c}-1} u^{T} _{k+i} R u_{k+i}
\end{equation}
where $N_p=8$ and $N_c=3$ represent the prediction and control horizons, and $U = [u^{T} _{k},...,u^{T} _{k+N_{c}-1}]^{T}$ is the vector of the input steps from sampling instant $k$ to sampling instant $k+N_{c}-1$. It was reported that prediction and control horizons are related to the speed of vehicles for a stable performance \citep{Keviczky2006}. Since a tractor is a slow vehicle, small prediction and control horizons are chosen to decrease computational burden in real-time applications.

The first sample of $U$ is applied to the plant:

\begin{equation}
u^{*}=u^{T} _{k}
\end{equation}
and the optimization problem is solved over a shifted horizon for the next sampling time. Q and R are positive-definite weighting matrices defined as follows:

\begin{equation}\label{}
Q  =  diag(0.5) \;\;\; , \;\;\; R  =  diag(1)
\end{equation}

The following plant objective function is solved at each sampling time for the MPC:

\begin{equation}
 \begin{aligned}
 & \underset{x(.), u(.)}{\text{min}}
 & & \displaystyle\sum\limits_{i=0}^{N_{p}} x^{T} _{k+i|k} Q x_{k+i|k} + \displaystyle\sum\limits_{i=0}^{N_{c}-1} u^{T} _{k+i} R u_{k+i}  \\
 & \text{subject to}
 && x(k+1)  =  A x(k) + B u(k)  \\
 && & y(k)  =  C x(k) + D u(k)\\
 &&& -45 \; degrees  \leq  \delta(k)  \leq 45  \;degrees  \\
 &&& -55 \; degrees/s  \leq  \Delta \delta(k)  \leq  55\; degrees/s
  \end{aligned}
  \label{lmpc}
\end{equation}

In our case, the designed MPC minimizes the error between the reference yaw rate and the measured yaw rate, and finds the desired steering angle $\delta_{desired}$ to the real-time system.

%%%%%%%%%%%%%%%%%%%%%%%%%%%%%%%%%%%%%%%%%%%%%%%%%%%%%%%%%%%%%%%%%%%%%%%%%%%%%%%%%%%%%%%%%%%%%%%%%%%%%%%%%%%%%%%%%%%%%%%%%%

\section{Trajectory Tracking Control}
\subsection{Overall Control Scheme}

\begin{figure*}[!ht]
\renewcommand{\arraystretch}{1.1}
\centering
  \includegraphics[width=6in]{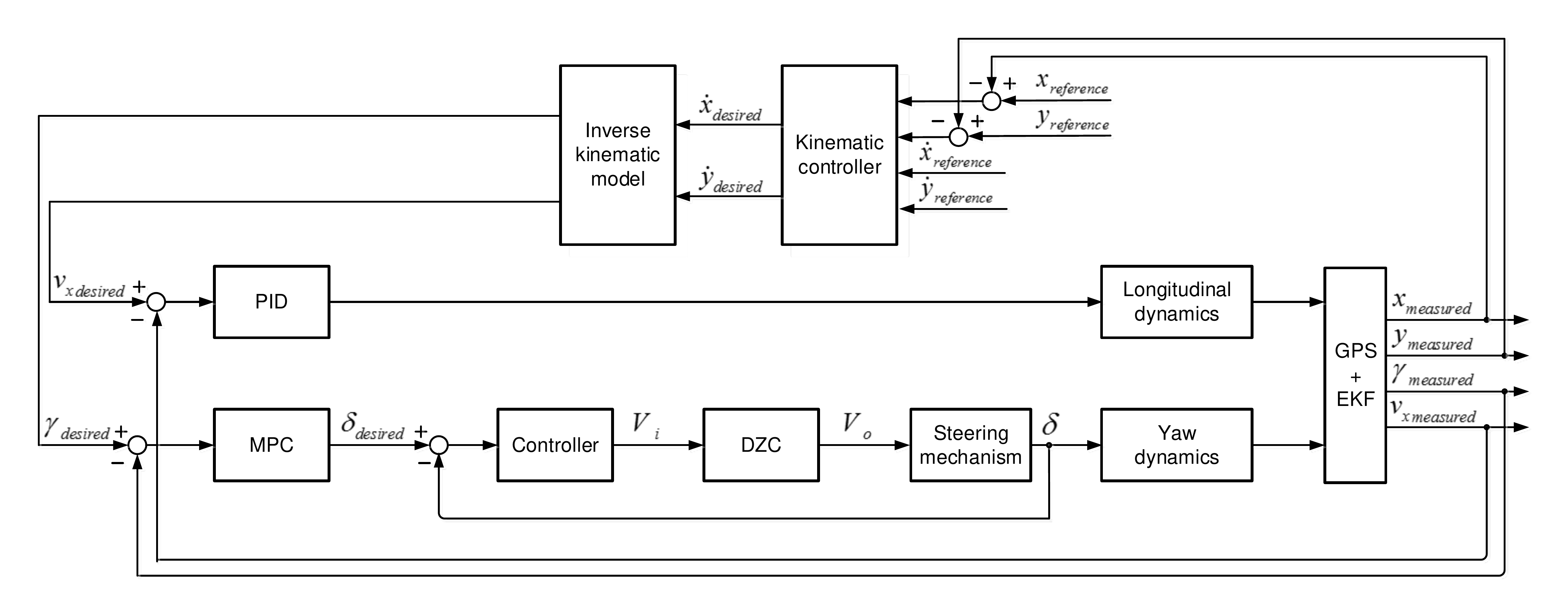}\\
  \caption{Block diagram of the proposed control scheme}
  \label{controlscheme}
\end{figure*}

\subsubsection{Kinematic Controller}
The kinematic model in (\ref{kinematicmodeltractor2}) is re-written in the algebraic model as follows:

\begin{eqnarray}\label{kinematicmodeltractorcontrol}
\left[
  \begin{array}{c}
   \dot{x} \\
   \dot{y} \\
  \end{array}
  \right]
  =
  \left[
  \begin{array}{cc}
  \cos{\psi} & - l_{r} \sin{\psi} \\
  \sin{\psi} &  l_{r} \cos{\psi} \\
  \end{array}
  \right]
    \left[
  \begin{array}{c}
   v_{x} \\
   \gamma \\
  \end{array}
  \right]
\end{eqnarray}
where the lateral velocity $v_{y}$ equals to $\gamma l_{r} $.

An inverse kinematic model is needed to generate the desired longitudinal speed and the desired yaw rate for the tractor. The inverse kinematic model in this study is written as follows:

\begin{eqnarray}\label{inversekinematicmodeltractor}
\left[
  \begin{array}{c}
  v_{x}  \\
  \gamma \\
  \end{array}
  \right]
  =
  \left[
  \begin{array}{cc}
  \cos{\psi}  & \sin{\psi}\\
  -\frac{1}{l_{r}} \sin{\psi} & \frac{1}{l_{r}} \cos{\psi} \\
  \end{array}
  \right]
    \left[
  \begin{array}{c}
   \dot{x} \\
   \dot{y}  \\
  \end{array}
  \right]
\end{eqnarray}

Considering $\dot{x}_{d} = \dot{x}_{r} + k_s \tanh{(k_c e_x)}$ and $\dot{y}_{d} = \dot{y}_{r} + k_s \tanh{(k_c e_y)}$, the kinematic control law proposed in \citep{Martins,erdalmech} to be applied to the tractor for the trajectory tracking control is written as:

\begin{eqnarray}\label{inversekinematicmodeltractor2}
\left[
  \begin{array}{c}
   v_{x_{d}} \\
  \gamma_{d} \\
  \end{array}
  \right]
  =
  \left[
  \begin{array}{cc}
  \cos{\psi}  & \sin{\psi}\\
  -\frac{1}{l_{r}} \sin{\psi} & \frac{1}{l_{r}} \cos{\psi} \\
  \end{array}
  \right]
    \left[
  \begin{array}{c}
   \dot{x}_{d} \\
   \dot{y}_{d} \\
  \end{array}
  \right]
\end{eqnarray}
where $v_{x_{d}}$ and $\gamma_{d}$ are respectively the desired speed and the desired yaw rate, and $e_x=x_r-x$ and $e_y=y_r-y$ are the current position errors in the $X-$ and $Y-$ axes, respectively. The parameters $k_{c}$ and $k_{s}$ are the gain and the saturation constant of the kinematic controller, respectively. The advantage of the kinematic model used in this paper is that since it has a saturation, the generated yaw rate cannot have extremely large values. The coordinates $(x,y)$ and $(x_r,y_r)$ are the current and the reference coordinates at the CG of the tractor, respectively.

Considering a perfect velocity tracking ($v_{x}=v_{x_{d}}$ and $\gamma=\gamma_{d}$) which means that the dynamics effects are ignored, the stability analysis is done by using a Lyapunov function \citep{Martins}.

\subsubsection{Dynamic Controllers}

The proposed control scheme used in this study is schematically illustrated in Fig. \ref{controlscheme}. A PID controller is used for the longitudinal velocity control. In the yaw dynamics control, an MPC controller is designed and its output is the desired steering angle for the front wheels. A low level PI controller is used to control the steering mechanism.

\subsubsection{State Estimation}

%Some states of the autonomous tractor cannot be measured (the yaw angle), while the measurements for some other states are delayed and noisy (x and y coordinate of the tractor, longitudinal speed). Moreover, the position data obtained from the RTK-GPS can be missing during several time steps when GPS or RTK-correction signal is not received.

An Extended Kalman Filter (EKF) was used for the state estimation. Since only one GPS antenna was mounted on the tractor, the yaw angle of the tractor cannot be measured. It is to be noted that the yaw angle of the tractor plays a very important role in the accuracy of trajectory tracking control as the estimated yaw angle is used in the inverse kinematic model to generate the desired speed and the desired yaw rate for the system. Therefore, the inputs of the EKF are position and velocity values from GPS. The outputs of the EKF are the position of the tractor on x- and y-coordinate system and the yaw angle. In trajectory control, the estimated values are used. Since the GPS antenna was located at point R on the tractor, the kinematic model in (\ref{kinematicmodeltractor}) is used. The discrete-time kinematic model used by EKF is written with a sampling interval $T_{s}$ as follows:

\begin{eqnarray}
\label{kinematicmodelEKF}
x_{k+1} & = & x_{k} + T_{s} v_{x_{k}} \cos{\psi_{k}} \nonumber \\
y_{k+1} & = & y_{k} + T_{s} v_{x_{k}} \sin{\psi_{k}} \nonumber \\
\psi_{k+1} & = & \psi_{k} + T_{s} v_{x_{k}} \frac{\tan{\delta_{k}}}{L}
\end{eqnarray}

The general form of the estimated system model is written:

\begin{eqnarray}
\label{generalmodelEKF}
\widehat{x}_{k+1} & = & f(\widehat{x}_{k}, u_{k}) + w_{k} \nonumber \\
\widehat{y}_{k+1} & = & h(\widehat{x}_{k}) + v_{k}
\end{eqnarray}
where $f$ is the estimation model for the system and $h$ is the measurement function. The difference between the kinematic model and estimation model is the process noise $w_{k}$ and observation noise $v_{k}$ both in the state and the measurement equations. They are both assumed to be independent and zero mean multivariate Gaussian noises with covariance matrices $Q_{k}$ and $R_{k}$, respectively:

\begin{eqnarray}
\label{noise}
w_{k} \backsim N(0,Q_{k}) \nonumber \\
v_{k} \backsim N(0,R_{k})
\end{eqnarray}

\subsection{Experimental Results}
\begin{figure}[t!]
\centering
  \includegraphics[width=3.4in]{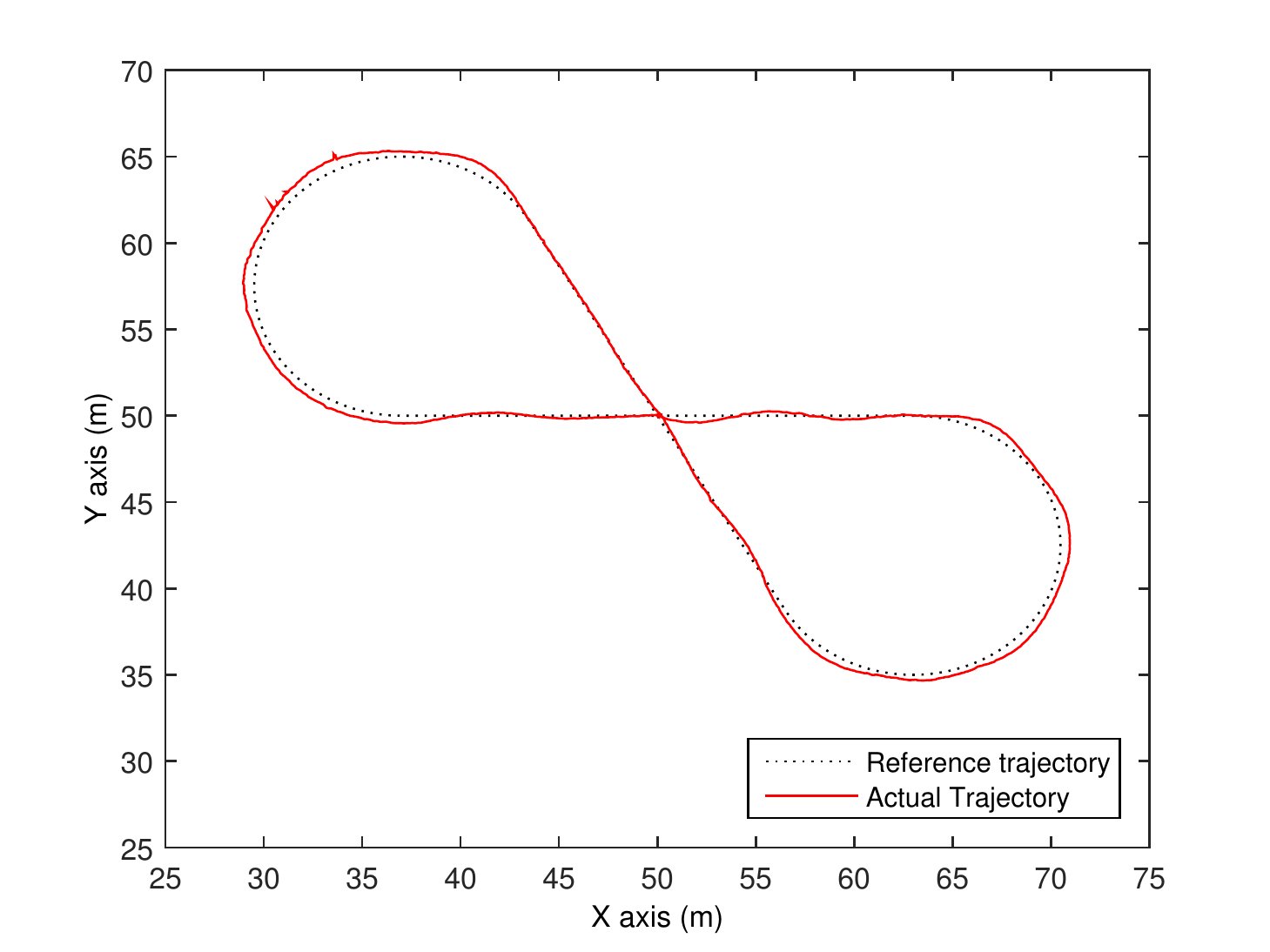}\\
  \caption{The reference and the actual trajectories of the autonomous tractor}
  \vskip -0.1cm
  \label{trajectory}
\end{figure}

\begin{figure}[t!]
\centering
  \includegraphics[width=3.4in]{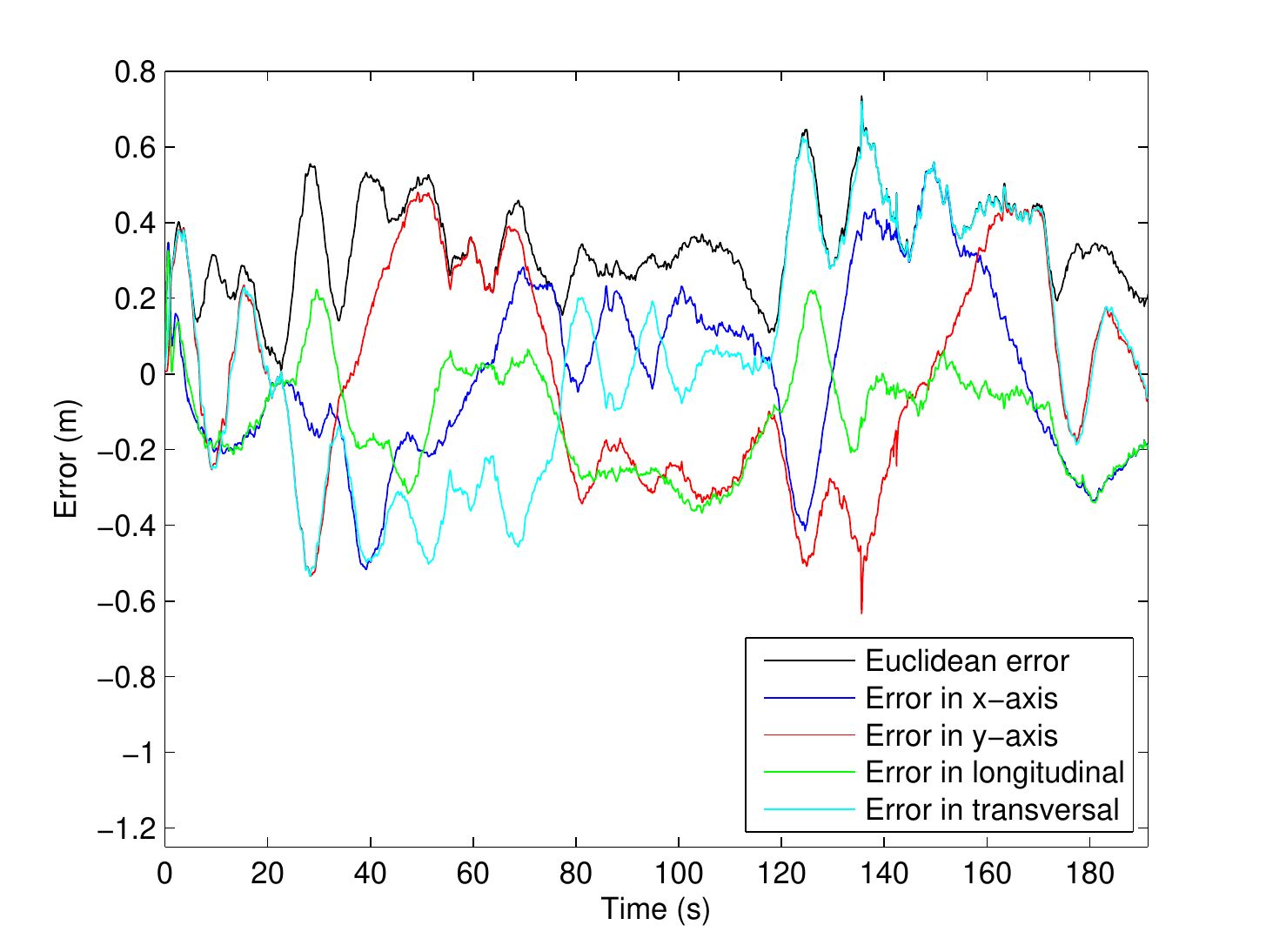}\\
  \caption{The trajectory following error both in x- and y-axes}
  \vskip -0.1cm
  \label{error}
\end{figure}

An 8-shaped trajectory with both straight and curved line geometries has been applied as the reference trajectory. The motivation of choosing an 8-shaped trajectory is that we can evaluate the performance of the controller both for straight and curved lines. The reference and actual trajectories of the autonomous tractor and the error values on the related trajectory are shown in Figs. \ref{trajectory} and \ref{error}, respectively. The experimental results show that the proposed control scheme is able to control the system.

Since the given trajectory consists of linear and curvilinear lines, the optimal values of the $Q$ and $R$ weighting matrices of the MPC controller had to be tuned by making a trade-off between optimal performance on the straight and curvilinear lines. It was observed during the experiments that although the curvilinear trajectory tracking is better with  aggressive controllers (having big gains) than the linear trajectory, these controllers give a large overshoot for the linear trajectory. Meanwhile, although the linear trajectory tracking is worse with an aggressive controller than the curvilinear trajectory, this controller gives large error values for the curvilinear trajectory. Thus, it can be concluded that if a reference signal consisting of only linear or curvilinear geometry is considered, more accurate results can be obtained.

\begin{figure}[t!]
\centering
  \includegraphics[width=3.4in]{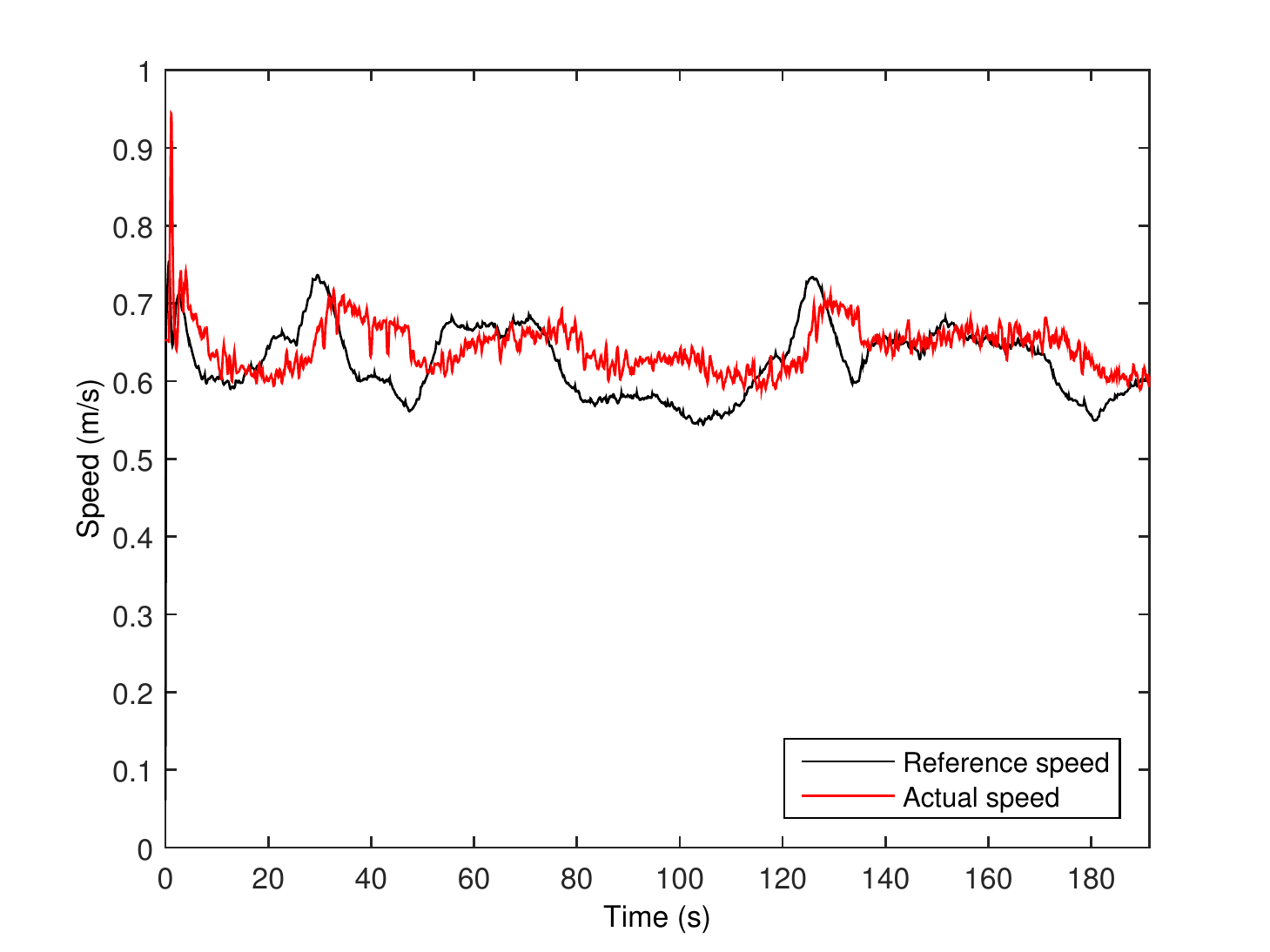}\\
  \caption{The reference and the actual longitudinal speed of the autonomous tractor}
  \vskip -0.1cm
  \label{speed}
\end{figure}

\begin{figure}[t!]
\centering
  \includegraphics[width=3.4in]{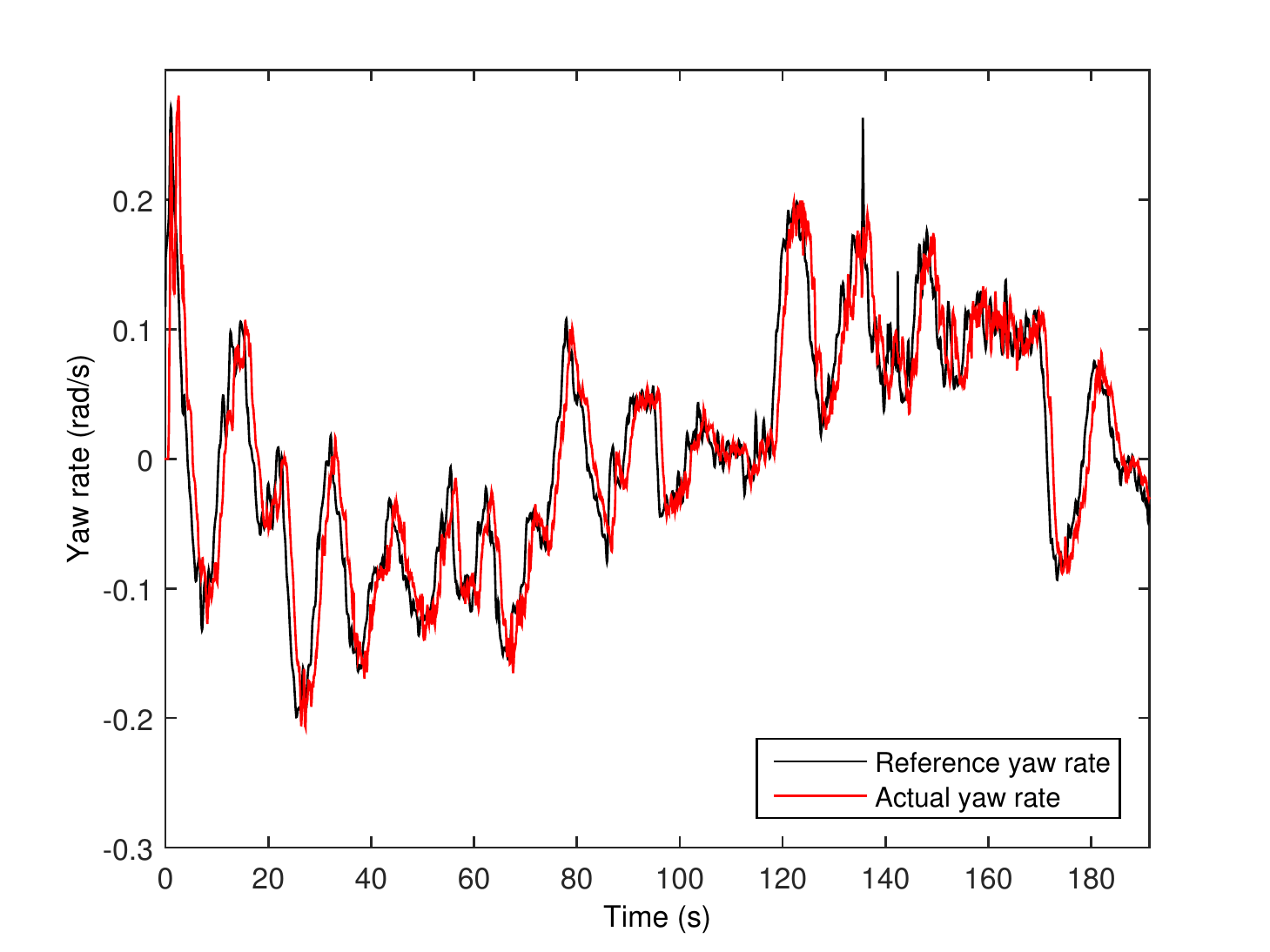}\\
  \caption{The reference and the actual yaw rate of the autonomous tractor}
  \vskip -0.1cm
  \label{yawrate}
\end{figure}

Figures \ref{speed} and \ref{yawrate} show the longitudinal velocity and the yaw rate responses of the autonomous tractor. As can be seen from Fig. \ref{speed}, there is a steady state error for the control of the longitudinal velocity by the PID controller. On the other hand, there is no steady state error for the yaw rate control by the MPC controller. These figures show that the error in both the x- and y-axes come from the poor control performance of the PID controller in the longitudinal dynamics, because it cannot cope with the strong high nonlinearities and interaction between the subsystems.


%%%%%%%%%%%%%%%%%%%%%%%%%%%%%%%%%%%%%%%%%%%%%%%%%%%%%%%%%%%%%%%%%%%%%%%%%%%%%%%%%%%%%%%%%%%%%%%%%%%%%%%%%%%%%%%%%%%%%%%%%%

\section{Conclusion}
In this study, modeling, identification and control aspects of an autonomous tractor have been investigated. Three yaw dynamics models have been derived from the equations of motion of the system. The parameters of these transfer functions have been estimated through nonlinear least squares frequency domain identification. While the fourth order transfer function model derived from the equations of motion considering the cornering stiffness and relaxation length for both the front and rear wheels gave realistic parameters estimates, the transfer function could be reduced to a second order model as not all yaw dynamics could be excited by the steering mechanism. This reduced order transfer function was then incorporated in an MPC for the yaw dynamics control and combined with a PID controller for the longitudinal speed control and an inverse kinematic controller for the trajectory tracking. The performance of these controllers was evaluated during real-time tests. The yaw rate control by the MPC gave satisfactory results, while the PID control of the longitudinal velocity did not. Although the 8-shaped time based reference trajectory could be tracked reasonably well, the longitudinal speed control should be improved to obtain better trajectory tracking. In order to increase the control accuracy in the straight lines, a space-based trajectory in which the longitudinal speed is constant, but only the yaw rate of the tractor is controlled, can be preferred. The MPC presented in this study provides the ideal framework for this.

% As in these real-time tests the side-slips did not vary, future research should focus on the incorporation of the estimated side-slips in the yaw dynamics control for the curvilinear trajectories.
%%%%%%%%%%%%%%%%%%%%%%%%%%%%%%%%%%%%%%%%%%%%%%%%%%%%%%%%%%%%%%%%%%%%%%%%%%%%%%%%%%%%%%%%%%%%%%%%%%%%%%%%%%%%%%%%%%%%%%%%%%

\appendix
\section{The parameters in the transfer functions}  \label{parameterstf}

The parameters in (\ref{BMwrltff}) and (\ref{BMwrltffr}) are written as follows:

\begin{eqnarray} \label{}
b_{0}^{\diamond} & = & \frac{C_{\alpha, f} C_{\alpha, r} (l_{f} + l_{r})}{I m \sigma _{f}}  \nonumber \\
b_{1}^{\diamond} & = & \frac{C_{\alpha, f} l_{f} v_{x}} {I \sigma _{f}} \nonumber \\
a_{0}^{\diamond} & = & \frac{m v_{x}^{2} ( - C_{\alpha, f} l_{f} + C_{\alpha, r} l_{r} ) + C_{\alpha, f} C_{\alpha, r} (l_{f} + l_{r})^{2} } {I m v_{x} \sigma _{f}}  \nonumber \\
a_{1}^{\diamond} & = & \frac{I (C_{\alpha, f} + C_{\alpha, r}) + m (C_{\alpha, f} l_{f} ^{2} + C_{\alpha, r} l_{r} ^{2}) + C_{\alpha, r} l_{r} m \sigma _{f} } {I m \sigma _{f}}  \nonumber \\
a_{2}^{\diamond} & = & I m v_{x} ^{2} + C_{\alpha, r} l_{r} ^{2} m \sigma _{f} + I C_{r} \sigma _{f} \nonumber \\
a_{3}^{\diamond} & = & 1 \nonumber \\
b_{0}^{\ast} & = & \frac{C_{\alpha, f} C_{\alpha, r} (l_{f} + l_{r}) v_{x}}{I m \sigma _{f} \sigma _{r}}  \nonumber \\
b_{1}^{\ast} & = & \frac{C_{\alpha, f} l_{f} v_{x} ^{2}} {I \sigma _{f} \sigma _{r}} \nonumber \\
b_{2}^{\ast} & = & \frac{C_{\alpha, f} l_{f} v_{x}} {I \sigma _{f}} \nonumber \\
a_{0}^{\ast} & = & \frac{ C_{\alpha, f} C_{\alpha, r} (l_{f} + l_{r})^{2} } {I m \sigma _{f} \sigma _{r}}  \nonumber \\
a_{1}^{\ast} & = & \frac{ I (C_{\alpha, f} + C_{\alpha, r}) + m v_{x} ( C_{\alpha, f} l_{f}^{2} + C_{\alpha, r} l_{r}^{2} - C_{\alpha, f} l_{f} \sigma_{r} + C_{\alpha, r} l_{r} \sigma_{f})} {I m \sigma _{f} \sigma _{r}}  \nonumber \\
a_{2}^{\ast} & = & \frac{I (m v_{x} ^{2} + C_{\alpha, f} \sigma_{r} + C_{\alpha, r} \sigma_{f} ) + m (C_{\alpha, f} l_{f} ^{2} \sigma_{r} + C_{\alpha, r} l_{r} ^{2} \sigma_{f})}{I m \sigma_{f} \sigma_{r}} \nonumber \\
a_{3}^{\ast} & = & \frac{(\sigma _{f} + \sigma _{r}) v_{x}}{\sigma_{f} \sigma_{r}} \nonumber \\
a_{4}^{\ast} & = & 1 \nonumber
\end{eqnarray}

%%%%%%%%%%%%%%%%%%%%%%%%%%%%%%%%%%%%%%%%%%%%%%%%%%%%%%%%%%%%%%%%%%%%%%%%%%%%%%%%%%%%%%%%%%%%%%%%%%%%%%%%%%%%%%%%%%%%%%%%%%

\section*{Acknowledgement}
This work has been carried out within the framework of the project IWT-SBO 80032 (LeCoPro) of the Institute for the Promotion of Innovation through Science and Technology in Flanders (IWT-Vlaanderen). We would like to thank Mr. Soner Akpinar for his technical support for the preparation of the experimental set up.

\bibliographystyle{elsarticle-harv}
\bibliography{JFR_ref}

\end{document}